\documentclass[12pt]{article}
\usepackage{amsmath}
\usepackage{amssymb}
\usepackage{epsfig}

\newcommand{\fr}[2]{\frac{#1}{#2}}
\newcommand{\del}{\partial}
\newcommand{\ba}{\begin{array}}
\newcommand{\ea}{\end{array}}
\newcommand{\bs}{\begin{split}}
\newcommand{\es}{\end{split}}
\newcommand{\be}{\begin{equation}}
\newcommand{\ene}{\end{equation}}
\newcommand{\bd}{\begin{displaymath}}
\newcommand{\ed}{\end{displaymath}}
\newcommand{\bea}{\begin{eqnarray}}
\newcommand{\eea}{\end{eqnarray}}
\newcommand{\beas}{\begin{eqnarray*}}
\newcommand{\eeas}{\end{eqnarray*}}
\newcommand{\bi}{\begin{itemize}}
\newcommand{\ei}{\end{itemize}}
\newcommand{\bt}{\begin{tabular}}
\newcommand{\et}{\end{tabular}}
\newcommand{\bc}{\begin{center}}
\newcommand{\ec}{\end{center}}
\newcommand{\bal}[1]{\renewcommand{\arraystretch}{#1}\begin{array}}
\newcommand{\eal}{\end{array}\renewcommand{\arraystretch}{1}}
\newcommand{\bals}{\begin{align*}}
\newcommand{\eals}{\end{align*}}

\newcounter{fig}
\renewcommand{\thefig}{\arabic{fig}}
\newcommand{\Caption}[2]{\refstepcounter{fig}\label{#1} Figure \thefig: #2}

\newtheorem{theorem}{\bf Theorem}[section]
\newtheorem{lemma}[theorem]{\bf Lemma}
\newtheorem{cor}[theorem]{\bf Corollary}

\newcommand{\Theorem}{\begin{theorem}\quad}
\newcommand{\Corollary}{\begin{cor}\quad}
\newcommand{\Lemma}{\begin{lemma}\quad}

\newcommand{\EndTheorem}{\end{theorem}}
\newcommand{\EndCorollary}{\end{cor}}
\newcommand{\EndLemma}{\end{lemma}}

\newcommand{\diverg}{\operatorname{div}}

\newcommand{\EL}{{\mathbb{L}}}

\newcommand{\Fou}{{\mathcal F}}

\newcommand{\ord}{{\mathcal O}}

\newcommand{\oh}{\frac{1}{2}}

\newcommand{\restr}[2]{\left.#1\right|_{{#2}} }
\newcommand{\du}{\left\langle}
\newcommand{\ud}{\right\rangle}

\newcommand{\eps}{\epsilon}

\newcommand{\delt}{{\triangle t}}
\newcommand{\delx}{{\triangle x}}

\newcommand{\Max}{{\mathcal M}}

\newcommand{\Ma}{\text{Ma}}
\newcommand{\Rey}{\text{Re}}
\newcommand{\bfx}{{\bf x}}
\newcommand{\bfv}{{\bf v}}
\newcommand{\bfu}{{\bf u}}
\newcommand{\bfn}{{\bf n}}

\title{A new discrete velocity method for Navier--Stokes equations}
\author{{\Large Michael Junk} \\ Department of Mathematics \\ 
  University of Kaiserslautern \\ D-67663, Germany
  \\[1mm]  and  \\[1mm]
  {\Large S.V.Raghurama Rao} \\ Institut f\"ur Techno-- und Wirtschaftsmathematik \\
  Erwin--Schr\"odinger-- Stra\ss e, Kaiserslautern \\ D-67663,
  Germany} \date{}
\begin{document}
\maketitle 
\begin{abstract}The relation between the Lattice Boltzmann Method, which has 
recently become popular, and the Kinetic Schemes, which are routinely 
used in Computational Fluid Dynamics, is explored.  A new discrete 
velocity model for the numerical solution of Navier--Stokes equations 
for incompressible fluid flow is presented by combining both the 
approaches.   
The new scheme can be interpreted as a pseudo-compressibility method
and, for a particular choice of parameters,
this interpretation carries over to the Lattice Boltzmann Method.
\end{abstract}

\section{Introduction}
In the last decade, Lattice Boltzmann Method has emerged as a
potential alternative to other Computational Fluid Dynamics techniques
in simulating fluid flows numerically.  The Lattice Boltzmann Method
(LBM) was first introduced by McNamara and Zanetti \cite{McNama1} to
overcome the drawbacks of the Lattice Gas Cellular Automata
(LGCA), which resulted from attempts to obtain macroscopic fluid
flow simulations from the simplest possible microscopic description
using discrete velocity models.
See the references \cite{Frisch,RothZal1} and \cite{RothZal2} for
reviews of the LGCA, references \cite{BenzSucciVerg} and \cite{ChenDool}
for reviews of the Lattice Boltzmann method and \cite{P&I88} for a review
on discrete velocity models.

\vspace{0.5\baselineskip}

Some authors noted the closeness of the Lattice Boltzmann Method to
the Kinetic Schemes (see \cite{HeLuo2,Cao}), which have been 
routinely used in CFD simulations
(see reference \cite{SMD4}
for a review of Kinetic Schemes).  
Both methods 
use the Boltzmann equation of Kinetic Theory as the starting
point, but are aimed at solving the macroscopic equations of fluid
flow. This approach exploits the
fact that the Boltzmann evolution is essentially equivalent to
Euler or Navier Stokes evolution if the state is in or close to
local thermodynamic equilibrium.
While most of the
Kinetic Schemes were developed for the solution of compressible
equations, the Lattice Boltzmann Method operates in the
incompressible limit.  
However, as we will show in Section \ref{s2},
the two methods even coincide for a particular parameter constellation.
This implies that the observations which are valid for Kinetic Schemes also have a direct
consequence on LBM.
In view of these remarks it is surprising that the close relation between 
the two methods is not fully appreciated. Our intention is to stress the remarkable coincidence.

\vspace{0.5\baselineskip}

The first Kinetic Scheme was 
introduced more than two decades back, by Sanders and Prendergast 
\cite{SandPrend}.  It is popularly known as the {\em Beam scheme}, 
and, incidentally, is also a discrete velocity model for simulating 
Euler equations.  
A few years later, an approach to construct Kinetic Schemes for general hyperbolic systems of
conservation laws was described by Harten, Lax and van Leer
\cite{HLL83}. Many Kinetic Schemes for the compressible Euler system based on the
original Maxwellian distribution were developed afterwards by
Pullin \cite{Pullin}, Reitz \cite{Reitz}, 
Deshpande and Mandal \cite{SMD2,SMD3,JCMSMD},
Perthame and Coron \cite{Pertham1,C&P91}, Prendergast and Xu \cite{PrendXu}, 
 Xu, Martinelli and Jameson \cite{XMJ95} and 
Raghurama Rao and Deshpande \cite{SVR2,SVR3}.
For the isentropic Euler system, Kaniel investigated a Kinetic Scheme based on an equilibrium distribution function which is different from the classical Maxwellian
\cite{Kan80,Kan88}.
A general approach to construct equilibrium distributions 
has been presented in \cite{Michael1} and by Perthame \cite{Pertham2}
who uses an entropy principle. For the compressible Navier-Stokes system,
Kinetic Schemes were developed by Chou
and Baganoff \cite{Baganoff} and in the group of Deshpande \cite{Manoj,Ramanan}.
A slightly different approach was taken by Xu and Prendergast \cite{XuPrend}.

For scalar conservation laws in one space dimension
B\"acker and Dressler found
equilibrium distributions following the idea of Kaniel \cite{BackerDressler}.
In arbitrary space dimensions the scalar case could be treated
with a slightly modified transport equation \cite{G&M83,P&T91}.
This approach led to a detailed investigation
of the relation between the hydrodynamic limit of kinetic equations 
and nonlinear conservation laws by Lions, Perthame and Tadmor \cite{LPT94}.

\vspace{0.5\baselineskip}

In this paper, we present a new discrete velocity model based
on the methodology of the Kinetic Schemes.
In Section \ref{s2}, the original concept of Kinetic Schemes for Euler
equations is introduced and then applied with a special equilibrium
distribution known from the Lattice Boltzmann Method. After that, the
obtained discrete velocity model is extended to the Navier Stokes case
by constructing a new discrete Chapman Enskog distribution.
In Section \ref{s4}, consistency of the resulting scheme is investigated,
leading to an interpretation of both Kinetic Schemes and LBM as
pseudo-compressibility methods.
Section \ref{s5} concludes with numerical results and discussions.

\section{Kinetic schemes for Euler equations} \label{s2}
\subsection{Traditional Kinetic Schemes in CFD}  
     The basis of Kinetic Schemes is the connection between the Boltzmann 
equation of Kinetic Theory of Gases and the macroscopic equations of 
fluid flow.   The fluid flow equations can be obtained as (velocity) 
moments of the Boltzmann equation  
\begin{equation} \label{BE}
\frac{\partial f}{\partial t} 
+ {\bf v} \cdot \nabla f=Q(f).
\end{equation}  
Here $f(\bfx,\bfv,t)$ is the velocity distribution function of the gas particles
and the gradient is taken with respect to the space variable $\bfx$.  
The left hand side of the equation (\ref{BE}) denotes the free flow 
of the molecules.  This free flow is disturbed by the molecular collisions, which 
is represented by the collision term, $Q(f)$, on the right hand side of the 
equation (\ref{BE}).  The mass, momentum and energy of the fluid 
can be obtained as the velocity averages of the particle mass, momentum 
and energy densities.    
Introducing the notation 
\begin{equation} \label{innerprod}
\du f \ud = \int_{- \infty}^{\infty} \int_{- \infty}^{\infty}\int_{- \infty}^{\infty}  
f \ dv_{1} \ dv_{2} \ dv_{3}  
\end{equation}
this can be formulated as 
\begin{equation} \label{moments1}  
\rho = \du f \ud \qquad \rho {\bf u} = \du {\bf v} f \ud \qquad \rho \epsilon  
= \du \fr{1} {2} |{\bf v}|^{2} f \ud 
\end{equation} 
The macroscopic equations can be obtained by integrating the Boltzmann 
equation (\ref{BE}), after multiplying it by the vector of the moment functions, as   
\begin{equation} \label{EE1}
\du \left( \ba{c} 1 \\ {\bf v} \\ \frac{1} {2} |{\bf v}|^{2} \ea \right)
\left( \fr{\del f} {\del t} + {\bf v} \cdot \nabla f -Q(f)\right) \ud = 0  
\end{equation}
Using (\ref{moments1}), we get the system 
\begin{equation}\label{EE2}
\begin{gathered}
 \frac{\partial \rho }{\partial t} 
+ \diverg\du {\bf v} f \ud =\du Q\ud\\
 \frac{\partial \left( \rho {\bf u} \right) }{\partial t} 
+ \diverg \du {\bf v}\otimes{\bf v} f\ud=\du {\bf v} Q\ud\\
 \frac{\partial \left(\rho \epsilon \right) } {\partial t} 
+ \diverg \du \frac{1} {2} |{\bf v}|^2 {\bf v} f\ud=\du \frac{1} {2} |{\bf v}|^2 Q\ud.
\end{gathered}
\end{equation}
The mass, momentum and energy are conserved during collisions.  Therefore, we have 
\begin{equation} \label{conscoll}
\du Q \ud = 0 \qquad \du {\bf v} Q \ud = 0 \qquad \du \fr{1} {2} |{\bf v}|^{2} Q \ud = 0 
\end{equation}
Substituting (\ref{conscoll}) in (\ref{EE2}) we obtain 
\begin{equation} \label{EE3}  
\begin{gathered}
 \frac{\partial \rho }{\partial t}
+ \diverg\du {\bf v} f \ud = 0 \\
 \frac{\partial \left( \rho {\bf u} \right) }{\partial t}
+ \diverg \du {\bf v}\otimes {\bf v} f\ud = 0 \\
 \frac{\partial \left( \rho \epsilon \right) }{\partial t}
+ \diverg \du \frac{1} {2} |{\bf v}|^2 {\bf v} f\ud = 0 
\end{gathered} 
\end{equation}
In the hydrodynamic limit, the gas is dominated by collisions and the 
particle distribution attains the form of a Maxwellian (in the limit of 
infinite collision frequency), given by 
\begin{equation} \label{Maxwel} 
\Max(v)=\frac{\rho}{(2\pi T)^\frac{3}{2}} 
\exp\left(-\frac{|{\bf v} - {\bf u}|^2}{2T}\right) 
\end{equation}
This velocity distribution is well known as the one of a gas in (local) 
thermodynamic equilibrium.  Hence, the Maxwellian is also called the {\em equilibrium 
distribution}.  When the distribution is a Maxwellian, the fluxes in (\ref{EE3}) 
can be calculated, yielding 
\begin{equation} \label{fluxes}
\du {\bf v}\otimes{\bf v} \Max\ud=\rho \bfu\otimes \bfu+\rho T I\quad\text{and} \quad 
\du \oh|{\bf v}|^2 {\bf v} \Max\ud=\rho(\epsilon+T)\bfu
\end{equation}
where $I$ is the identity matrix.  Using the above, we obtain the Euler equations as 
\begin{equation} \label{EE4}
\begin{gathered}
\frac{\partial \rho}{\partial t} + \diverg (\rho {\bf u})=0 \\
\frac{\partial (\rho {\bf u})}{\partial t} + \diverg (\rho {\bf u}\otimes {\bf u} +\rho T I)= 0 \\
\frac{\partial (\rho \epsilon)}{\partial t} + \diverg (\rho (\epsilon + T){\bf u}) = 0
\end{gathered} 
\end{equation} 
or equivalently in the form  
\begin{equation} \label{EE5} 
\du \left( \ba{c} 1 \\ {\bf v} \\ \frac{1} {2} |{\bf v}|^{2} \ea \right) 
\left( \fr{\del f} {\del t} + {\bf v} \cdot \nabla f\right) \ud = 0, \qquad f = \Max  
\end{equation} 
While standard discretizations of the Euler system are based on \eqref{EE4}, Kinetic
Schemes use the representation \eqref{EE5} which is motivated by Kinetic Theory.
An obvious advantage of \eqref{EE5} is the much simpler differential operator
which is linear and scalar in contrast to the more complicated nonlinear system \eqref{EE4}.

To discretize \eqref{EE5}, traditional CFD
techniques 
like finite difference, finite volume, finite element or spectral 
methods can be applied.
Equivalently, one can use
the {\em Lagrangian approach}.  
In this approach, we replace \eqref{EE5} by the auxiliary problem
\begin{equation} \label{linearBE}
\fr{\del f}{\del t} + {\bf v} \cdot \nabla f = 0, \qquad \restr{f}{t=0} = \Max
\end{equation} 
for which the solution is straightforward, given by 
\begin{equation} \label{linearsol}
f({\bf x}, {\bf v}, t)=f({\bf x} - {\bf v}t, {\bf v}, 0)
\end{equation}
Clearly, this solution satisfies 
\begin{equation*} 
\du \left( \ba{c} 1 \\ {\bf v} \\ \frac{1} {2} |{\bf v}|^{2} \ea \right)
\left( \fr{\del f} {\del t} + {\bf v} \cdot \nabla f\right) \ud = 0.
\end{equation*}
However, the constraint $f=\Max$ is enforced only initially.  The 
violation of this constraint leads to an increasing error as time 
increases.  By stopping the evolution after a small time step 
$\delt$ and restarting it with a Maxwellian (that has the same 
$\rho,u, \epsilon$--moments as the solution of the just finished free 
flow step), the error can be kept of order $\delt$, giving rise 
to a first order method for the Euler equations.  Thus, two clear steps 
can be identified for the Lagrangian approach: a convection step and 
a relaxation step.  In the relaxation step, the velocity distribution 
relaxes completely to the equilibrium distribution.

\subsection{Kinetic Schemes with discrete distributions} 
    While the Kinetic Schemes mentioned in the above subsection are 
designed to solve the Euler system, it is not necessary to be limited by this
restriction. Also, the choice of $\Max$ as equilibrium constraint is not
mandatory. Obviously, the approach is applicable whenever the system of equations allows
a representation of type \eqref{EE5}. In the following, we are going to restrict our
considerations to the case of isothermal Euler equations, in order to work out the
similarities with the Lattice Boltzmann method. The isothermal equations
(with $T=T_0=c_s^2$ where $c_s$ is the sound speed) are of the form
\begin{equation} \label{EE6}
\begin{gathered}
\frac{\partial \rho}{\partial t} + \diverg (\rho {\bf u})=0 \\
\frac{\partial (\rho {\bf u})}{\partial t} + \diverg (\rho {\bf u}\otimes {\bf u} +c_s^2\rho I)= 0
\end{gathered} 
\end{equation} 
or equivalently 
\begin{equation} \label{EE7} 
\du \left( \ba{c} 1 \\ {\bf v} \ea \right) 
\left( \fr{\del f} {\del t} + {\bf v} \cdot \nabla f\right) \ud = 0, \qquad f = \restr{\Max}{T=T_0}.  
\end{equation} 
Instead of the classical Maxwellian $\Max$ (with fixed temperature $T=T_0$), we can also
choose any distribution $F$, as long as the integral expressions in \eqref{EE7}
together with the constraint $f=F$ is equivalent to the Euler system \eqref{EE6}. Since the
integral involves velocity moments up to second order, we are led to the following
compatibility conditions on the equilibrium distribution
\begin{equation} \label{moments2}
\begin{gathered} 
\du F \ud = \rho \\ 
\du {\bf v} F \ud = \rho {\bf u} \\  
\du {\bf v} \otimes {\bf v} F \ud=\rho {\bf u}\otimes {\bf u} + c_s^2\rho I
\end{gathered}
\end{equation}
In particular, we are interested in discrete velocity distributions $F$ 
which satisfy these moment constraints
(see also the works of Sanders \& Prendergast
\cite{SandPrend}, Nadiga \& Pullin \cite{NadigaPullin} and Michael 
Junk (\cite{Michael2,Michael3}). An explicit example in 2D
is given by the so called D2Q9 distribution used in the Lattice Boltzmann method.
It is of the form
\begin{equation} \label{discretedist}
F(\rho,{\bf u};{\bf v}) = \sum_{i=0}^{8} F_{i}(\rho,{\bf u}) \ \delta \left( {\bf v} - {\bf v}_{i} \right)
\end{equation}
where $\delta$ is the Dirac-Delta function and  
\begin{equation} \label{discretevel}
\begin{array}{l}
{\bf v}_{0} = 0 \\ 
{\bf v}_{i} = \sqrt{3} c_s\left( 
\cos \left( \ \left(i - 1\right) \fr{\pi} {2} \right), \  
\sin \left( \left(i - 1\right) \fr{\pi} {2} \right) \  
\right)^{T} \hfill i = 1, \ldots, 4 \\ 
{\bf v}_{i} = \sqrt{6} c_s\left( 
\cos \left( \ \left(i - \fr{9} {2}\right) \fr{\pi} {2} \right), \  
\sin \left( \left(i - \fr{9} {2}\right) \fr{\pi} {2} \right) \  
\right)^{T} \hfill i = 5, \ldots, 8 
\end{array}
\end{equation}
The weights $F_i$ are given by
\begin{equation} \label{discretedist2} 
F_{i}(\rho,{\bf u}) = F_{i}^{*}\rho \left( 1 - \fr{1} {2c_s^2} |{\bf u}|^{2} + \frac{1}{c_s^2}{\bf u} \cdot {\bf v}_{i}
+ \fr{1} {2c_s^4} \left( {\bf u} \cdot {\bf v}_{i} \right)^{2} \right) 
\end{equation}
with 
\begin{equation} \label{discretewts}
F^{*}_{0} = \fr{4} {9} \qquad 
F^{*}_{i} = \fr{1} {9} \ {\rm for} \ i=1,\ldots,4 \qquad 
F^{*}_{i} = \fr{1} {36} \ {\rm for} \ i=5,\ldots,8 \qquad 
\end{equation}
In order to obtain a Kinetic Scheme for the isothermal Euler equations,
we will approximate the equivalent form 
\begin{equation} \label{EE8} 
\du \left( \ba{c} 1 \\ {\bf v} \ea \right) 
\left( \fr{\del f} {\del t} + {\bf v} \cdot \nabla f\right) \ud = 0, \qquad f =F
\end{equation} 
with the Lagrangian approach described in the last section. Solving the free flow
equation $\fr{\del f}{\del t}+{\bf v}\cdot\nabla_x f=0$, starting at time $t$  with equilibrium
$f({\bf x},{\bf v},t)=F(\rho({\bf x},t),{\bf u}({\bf x},t);{\bf v})$, yields after a time step $\delt$
\begin{align*}
f({\bf x},{\bf v},t+\delt)&= F(\rho({\bf x}-{\bf v}\delt,t),{\bf u}({\bf x}-{\bf v}\delt,t);{\bf v})\\
&=\sum_{i=0}^8 F_i(\rho({\bf x}-{\bf v}\delt,t),{\bf u}({\bf x}-{\bf v}\delt,t))\delta({\bf v}-{\bf v}_i).
\end{align*}
Using the relation $\psi({\bf v})\delta({\bf v}-{\bf v}_i)=\psi({\bf v}_i)\delta({\bf v}-{\bf v}_i)$,
which holds for any continuous function $\psi$, we obtain further
\begin{equation*}
f({\bf x},{\bf v},t+\delt)=\sum_{i=0}^8 F_i(\rho({\bf x}-{\bf v}_i\delt,t),{\bf u}({\bf x}-{\bf v}_i\delt,t))\delta({\bf v}-{\bf v}_i).
\end{equation*}
Denoting the weights of the discrete distribution $f({\bf x},{\bf v},t+\delt)$ by $f_i({\bf x},t+\delt)$,
the evolution can also be described without mentioning the Dirac deltas at the (fixed) discrete velocities 
\begin{equation} \label{LBKS}
f_i({\bf x},t+\delt)=F_i(\rho({\bf x}-{\bf v}_i\delt,t),{\bf u}({\bf x}-{\bf v}_i\delt,t)),\qquad i=0,\dots,8.
\end{equation}
Since integer multiples of $\bfv_i\delt$ make up a regular square grid which
is invariant under $\bfv_i\delt$--translations, the scheme \eqref{LBKS}
only accesses nodal data if $\bfx$ is also a node of the grid.
We remark that the grid length is given by
\begin{equation}\label{gridl}
\delx=\sqrt{3}c_s\delt.
\end{equation}
The connection between \eqref{LBKS} and the classical Lattice Boltzmann method 
becomes most obvious under the change  of
variables ${\bf x}\mapsto {\bf x}+{\bf v}_i\delt$, which leads to 
\begin{equation}\label{LBKS3}
f_i({\bf x}+{\bf v}_i\delt,t+\delt)=F_i(\rho({\bf x},t),{\bf u}({\bf x},t)),\qquad i=0,\dots,8.
\end{equation}
Indeed, \eqref{LBKS3} coincides with the 
Lattice Boltzmann evolution \cite{ChenChenMattheus,Qian}
\begin{equation} \label{LBE}
f_{i} \left( {\bf x} + {\bf v}_{i} \Delta t, t + \Delta t \right) 
- f_{i} \left( {\bf x}, t \right) 
= \fr{\delt} {t_{R}} \left( F_{i} \left({\bf x}, t\right) -  f_{i} \left({\bf x}, t\right) \right)  
\end{equation}
if we set $t_{R} = \delt$ (see references \cite{ChenChenMattheus} and 
\cite{Qian} for the LBM based on BGK-model).     
At first glance, this seems to be a contradiction, because the
Kinetic Scheme has been set up for the Euler system while it is known that the Lattice
Boltzmann method approximates the Navier Stokes system. In fact, setting
$t_R=\delt$ amounts to a high dose of viscosity (typically, LBM applications are run with 
$t_R$ in-between $\oh\delt$ and $\delt$). The apparent contradiction is resolved with the
remark that the Lagrangian approach to Kinetic Schemes yields only a first order
method. The numerical viscosity of that scheme is quite high, particularly in applications
with low Mach number flows. This numerical viscosity of the Kinetic Scheme is exactly
the viscosity corresponding to $t_R=\delt$ in LBM and thus has a physically correct structure.
In \cite{Michael3}, a Kinetic Scheme for the Euler system could therefore be used as solver for the
Navier Stokes equations.
Huang {\em et al.} \cite{HuangQian}, in their Lattice Boltzmann Method for 
compressible flows, used a similar approach, even though they did not mention 
Kinetic Schemes.  

\section{Extension to Navier--Stokes equations} \label{s3}
     Kinetic Schemes can also be extended to the case of 
Navier--Stokes equations, by using a Chapman--Enskog distribution 
function $\Fou_{CE}$ instead of the Maxwellian constraint $\Max$ in \eqref{EE6}.
This approach has been pursued in \cite{Baganoff,Manoj,Ramanan}.
The Chapman--Enskog distribution function $\Fou_{CE}$ 
is obtained as a small perturbation of the Maxwellian.  
See references \cite{Vincenti,Huang,SMD2}
and \cite{Kogan}  for details of the derivation.
For a mono-atomic gas in three dimensions, the distribution function is of the form
\begin{equation} \label{fCE}
\Fou_{CE} = \Max \left[ 
1 - \fr{P_{ij}} {p} \fr{1} {2 T} c_{i} c_{j} 
- \fr{q_{i}} {p} \fr{1} {T} c_{i} \left(1 - \fr{2} {5} \fr{c^{2}} {2 T} \right)   
\right]  
\end{equation}
where 
\begin{equation} \label{fCEdefs} 
q_{i} = - K \fr{\del T} {\del x_{i}}, \qquad
P_{ij} = - \mu \left(
\fr{\del u_{i}} {\del x_{j}} + \fr{\del u_{j}} {\del x_{i}} 
- \fr{2} {3} \delta_{ij} \fr{\del u_{k}} {\del x_{k}}  
\right) 
\end{equation}
and ${\bf c} = {\bf v} - {\bf u}$ is the peculiar velocity.
Here, we will again consider the simpler case of isothermal equations in 2D. Following \cite{HouZou,Michael2},
the equations are of the form 
\begin{equation}\label{NS1}
\begin{gathered}
 \frac{\partial \rho}{\partial t}+\diverg(\rho {\bf u})=0,\\
 \frac{\partial (\rho{\bf u}) }{\partial t}+\diverg(\rho {\bf u}\otimes{\bf u} +c_s^2\rho I)=
\diverg \eta
\end{gathered}
\end{equation}
where the viscous stress tensor is given by 
\begin{equation*}
\eta=\nu\rho(2S+\diverg{\bf u}I),\qquad
S_{ij}=\oh\left(\frac{\partial u_i}{\partial x_j}+\frac{\partial u_i}{\partial x_j}\right).
\end{equation*}
(Note that $\diverg\eta$ is the vector obtained by applying divergence to the rows of $\eta$.)
Equivalently, we can write system \eqref{NS1} as
\begin{equation} \label{NS2} 
\du \left( \ba{c} 1 \\ {\bf v} \ea \right) 
\left( \fr{\del f} {\del t} + {\bf v} \cdot \nabla f\right) \ud = 0, \qquad f = F_{CE}
\end{equation} 
where $F_{CE}$ satisfies the moment constraints
\begin{equation} \label{moments3}
\begin{gathered} 
\du F_{CE} \ud = \rho \\ 
\du  {\bf v} F_{CE} \ud = \rho  {\bf u} \\  
\du {\bf v}\otimes{\bf v} F_{CE} \ud=\rho {\bf u}\otimes{\bf u}+c_s^2\rho I-\eta
\end{gathered}
\end{equation}
Instead of using the continuous distribution function $\Fou_{CE}$ (given in \eqref{fCE}
with $q_i=0$ and $T=T_0$), we construct 
a discrete $F_{CE}$ satisfying \eqref{moments3}.
One possibility is to discretize the classical Chapman--Enskog distribution function in the
velocity variable. A Kinetic Scheme based on this approach will be presented elsewhere \cite{SVRMJ}.
Here, we follow a different idea based on a general solution technique for moment problems of the form
\eqref{moments3} which uses orthogonal polynomials \cite{Michael1}. For the special D2Q9 model,
however, it is not necessary to work out the general ideas. In fact, conditions \eqref{moments3}
can be reduced to those of the Euler system if we replace $\rho{\bf u}\otimes{\bf u}$ by
$\rho{\bf u}\otimes{\bf u}-\eta$.
This observation can be used, if we write the weights \eqref{discretedist2} of the equilibrium distribution function \eqref{discretedist}
in terms of $\rho{\bf u}\otimes{\bf u}$. Introducing the matrix product
\begin{equation*}
A:B=\sum_{i,j=1}^2 A_{ij}B_{ij}
\end{equation*}
we find
\begin{equation*}
{\bf u}\otimes{\bf u}:{\bf v}\otimes{\bf v}=\sum_{i,j=1}^2 u_iu_jv_iv_j=({\bf u}\cdot{\bf v})^2
\end{equation*}
and
\begin{equation*}
{\bf u}\otimes{\bf u}:I=\sum_{i,j=1}^2 u_iu_j\delta_{ij}=|{\bf u}|^2
\end{equation*}
so that \eqref{discretedist2} can be written as 
\begin{equation*}
F_{i}(\rho,{\bf u}) = F_{i}^{*}\rho \left( 1 +\frac{1}{c_s^2}{\bf u} \cdot {\bf v}_{i}+ \fr{1} {2c_s^2} {\bf u}\otimes{\bf u}:\left(\frac{1}{c_s^2}{\bf v}_i\otimes{\bf v}_i-I\right)\right).
\end{equation*}
Replacing ${\bf u}\otimes{\bf u}$ by ${\bf u}\otimes{\bf u}-\nu (2S+\diverg{\bf u}I)$, we finally obtain
\begin{multline}\label{nf}
F_{CE,i}(\rho,{\bf u})\\
= F_{i}^{*}\rho \left( 1 +\frac{1}{c_s^2}{\bf u} \cdot {\bf v}_{i}+ 
\fr{1} {2c_s^2} ({\bf u}\otimes{\bf u}-2\nu S-\nu\diverg{\bf u}I):\left(\frac{1}{c_s^2}{\bf v}_i\otimes{\bf v}_i-I\right)\right),
\end{multline}
or after going back to scalar products in ${\bf v}_i$ and ${\bf u}$,
\begin{multline} \label{discretefce1}
F_{CE,i}(\rho,{\bf u}) = F_{i}^{*}\rho \biggl( 1 +\frac{1}{c_s^2}{\bf u} \cdot {\bf v}_{i}
-\frac{1}{2c_s^2}|{\bf u}|^2+\frac{1}{2c_s^4}({\bf u}\cdot{\bf v}_{i})^2\\
-\frac{\nu}{c_s^4} S:{\bf v}_i\otimes{\bf v}_i-\frac{\nu}{c_s^2}\diverg{\bf u}\left(\frac{1}{2c_s^2}|{\bf v}_i|^2-2\right)\biggr).
\end{multline}
It is easy to check that the so defined Chapman--Enskog distribution
\begin{equation} \label{discretefce2}
F_{CE}(\rho,{\bf u};{\bf v})=\sum_{i=0}^8 F_{CE,i}(\rho,{\bf u}) \delta({\bf v}-{\bf v}_i)
\end{equation}
satisfies \eqref{moments3}. We also remark that $F_{CE}$ is a perturbation of the original D2Q9 equilibrium distribution,
similar to the classical case in Kinetic Theory where \eqref{fCE} is a perturbation of the Maxwellian \eqref{Maxwel}.

To develop a Kinetic Scheme for Navier--Stokes equations, we follow 
the same procedure as in the previous section, except that the 
distribution used as constraint after every time--step will now be 
the Chapman--Enskog distribution, $F_{CE}$. We end up with the scheme
\begin{equation} \label{LBKS2}
f_i({\bf x},t+\delt)=F_{CE,i}(\rho({\bf x}-{\bf v}_i\delt,t),{\bf u}({\bf x}-{\bf v}_i\delt,t)),\qquad i=0,\dots,8
\end{equation}
where the moments are updated according to
\begin{equation} \label{momupdate}
\begin{gathered}
\rho({\bf x},t+\delt)=\sum_{i=0}^8 f_i({\bf x},t+\delt)\\
 {\bf u}({\bf x},t+\delt)=\fr{1}{\rho({\bf x},t+\delt)}\sum_{i=0}^8{\bf v}_if_i({\bf x},t+\delt).
\end{gathered}
\end{equation}
 
\section{The incompressible limit} \label{s4}
To investigate the behavior of the Kinetic Scheme at low Mach numbers,
we first scale the compressible Navier Stokes system appropriately.
Low Mach number flows appear if ${\bf u}$
is very small compared to $c_s$. Taking a typical speed $U$ and length scale $L$
of the flow, the time scale $\Theta$ is chosen in accordance to these scales as
\begin{equation*}
\Theta=\frac{L}{U}.
\end{equation*}
The density $\rho$ is assumed to be of order one so that no scaling is needed.
To avoid superscripts, we will not change the symbols for scaled
functions and variables. If we refer to unscaled quantities (which
appear less often in this section), we add a hat to the symbols.
After some algebra, we obtain the scaled version of \eqref{NS1} 
\begin{equation*}
\begin{gathered}
 \frac{\partial \rho}{\partial t}+\frac{\Theta U}{L}\diverg(\rho {\bf u})=0,\\
 \frac{\partial (\rho{\bf u}) }{\partial t}+\frac{\Theta U}{L}\diverg(\rho {\bf u}\otimes{\bf u}) +\frac{c_s^2\Theta }{LU}\nabla \rho=
\frac{\Theta}{L^2}\diverg \eta
\end{gathered}
\end{equation*}
By assumption, $\Theta U/L=1$ and $c_s^2\Theta/(LU)=c_s^2/U^2$.
Introducing the Mach number $\Ma=U/c_s$ and the Reynolds number $\Rey=UL/\nu$
of the flow, we end up with
\begin{equation}\label{NSs}
\begin{gathered}
 \frac{\partial \rho}{\partial t}+\diverg(\rho {\bf u})=0,\\
 \frac{\partial (\rho{\bf u}) }{\partial t}+\diverg(\rho {\bf u}\otimes{\bf u}) +\frac{1}{\Ma^2}\nabla \rho=
\frac{1}{\Rey}\diverg (2\rho S+\rho\diverg{\bf u}I)
\end{gathered}
\end{equation}
If we approximate \eqref{NSs} by the Kinetic Scheme introduced in \eqref{LBKS2}, then the time step is related to
the grid length by $\Delta \hat{t}=\sqrt{3}c_s\Delta \hat{x}$ (see \ref{gridl}), or in scaled quantities
\begin{equation} \label{CFL}
\delt=\sqrt{3}\Ma\delx.
\end{equation}
This relation already indicates the typical problem that any explicit solver for the compressible equations faces
in the incompressible limit: to get a reasonable space resolution, the time resolution must be
extremely fine (if $\Ma\ll1$) to satisfy the CFL--condition \eqref{CFL}. 

To find out which equations are approximated by the Kinetic Scheme,
we perform a consistency analysis in the coupled limit $\delt,\Ma\to0$. More precisely, we assume
\begin{equation}\label{coup}
\frac{\delt}{\Ma^2}=\lambda=\text{const} \qquad \text{for $\delt\to0$, $\Ma\to0$}.
\end{equation}
To begin with, let us rewrite the Chapman Enskog distribution \eqref{nf}
in scaled quantities.
\begin{multline}\label{wgts}
F_{CE,i}(\rho,{\bf u})
= F_{i}^{*}\rho \Biggl( 1 +\Ma^2\,{\bf u} \cdot {\bf v}_{i}\\
+\fr{\Ma^2} {2} \left({\bf u}\otimes{\bf u}-\frac{1}{\Rey}(2 S+\diverg{\bf u}I)\right):(\Ma^2{\bf v}_i\otimes{\bf v}_i-I)\Biggr).
\end{multline}
Setting $F_{CE}(\rho,\bfu;\bfv)=\sum_{i=0}^8 F_{CE,i}(\rho,{\bf u})\delta(\bfv-\bfv_i)$ and using $\rho(\bfx)$ and $\bfu(\bfx)$ as initial values, the Kinetic Scheme
yields at the end of the first time step
\begin{equation}\label{req}
\rho^1(\bfx)=\du F_{CE}(\rho(\bfx-\bfv\delt),{\bf u}(\bfx-\bfv\delt);{\bfv})\ud
\end{equation}
and
\begin{equation}\label{req2}
(\rho^1\bfu^1)(\bfx)=\du {\bfv} F_{CE}(\rho(\bfx-\bfv\delt),{\bf u}(\bfx-\bfv\delt);{\bfv})\ud.
\end{equation}
To obtain a Taylor expansion around $\delt=0$ we need $\delt$--derivatives of \eqref{req} and \eqref{req2} up to a certain order. Obviously,
each $\delt$--derivative leads to a space derivative with $-\bfv$ as factor
(i.e. $\fr{\del}{\del\delt}=-v_i\fr{\del}{\del x_i}$). To get first order consistency 
in $\delt$, we nevertheless need higher $\delt$--derivatives. This is due to the fact that terms of the form
$\delt^2/\Ma^2$, $\delt^2/\Ma^3$ and $\delt^3/\Ma^4$ are not negligible
in the coupled limit \eqref{coup}.
Consequently, we also need higher order $\bfv$--moments of the Chapman Enskog distribution.
Taking the scaling into account, we get from
\eqref{moments3}
\begin{equation} \label{mom0-2}
\begin{gathered}
\du F_{CE} \ud = \rho \\
\du  {\bf v} F_{CE} \ud = \rho  {\bf u} \\
\du {\bf v}\otimes{\bf v} F_{CE} \ud=\rho {\bf u}\otimes{\bf u}+\frac{1}{\Ma^2}\rho I-\frac{1}{\Rey}(2\rho S+\rho\diverg\bfu I).
\end{gathered}
\end{equation}
The third order moment can be calculated using the explicit form of $F_{CE}$
given in \eqref{wgts}. We find
\begin{equation}\label{mom3}
\du v_iv_jv_k F_{CE}\ud=\frac{1}{\Ma^2}\rho(\delta_{ij}u_k+\delta_{ik}u_j+\delta_{kj}u_i).
\end{equation}
Finally, from the fourth and fifth order moments we only need to know the terms of leading order
\begin{equation}\label{mom4-5}
\begin{gathered}
\du v_iv_jv_kv_lF_{CE}\ud=
\frac{1}{\Ma^4}\rho(\delta_{ij}\delta_{kl}+\delta_{ik}\delta_{jl}+\delta_{il}\delta_{jk})+\ord\left(\frac{1}{\Ma^2}\right)\\
\du v_iv_jv_kv_lv_mF_{CE}\ud=
\ord\left(\frac{1}{\Ma^4}\right).
\end{gathered}
\end{equation}
The Taylor expansion of \eqref{req} is then given by
\begin{multline*}
\rho^1 = \du F_{CE}\ud - \frac{\del}{\del x_i}\du v_i F_{CE}\ud{\delt}
 +\oh \frac{\del^2}{\del x_i\del x_j}\du v_iv_j F_{CE}\ud\delt^2\\
 -\frac{1}{6} \frac{\del^3}{\del x_i\del x_j\del x_k}\du v_iv_jv_k F_{CE}\ud
\delt^3+\ldots
\end{multline*}
Since
\begin{equation}\label{himo}
\begin{gathered}
\du v_iv_j  F_{CE} \ud\delt^2=\frac{\delt^2}{\Ma^2}\rho\delta_{ij}+\ord(\delt^2)\\
\du v_iv_jv_k  F_{CE} \ud\delt^3=\ord\left(\frac{\delt^3}{\Ma^2}\right)=\ord(\delt^2)\\
\du v_iv_jv_kv_l  F_{CE} \ud\delt^4=\ord\left(\frac{\delt^4}{\Ma^4}\right)=\ord(\delt^2)
\end{gathered}
\end{equation}
we conclude
\begin{equation}\label{tdensity}
\rho^1=\rho+\left(\oh\lambda\Delta\rho-\diverg(\rho\bfu)\right)\delt +\ord(\delt^2).
\end{equation}
Similarly, we get for the momentum defined in \eqref{req2}
\begin{multline*}
(\rho^1\bfu^1)_l = \du v_l F_{CE}\ud
 - \frac{\del}{\del x_i}\du v_i v_l F_{CE}\ud\delt
 +\oh \frac{\del^2}{\del x_i\del x_j}\du v_iv_j v_l F_{CE}\ud\delt^2\\
 -\frac{1}{6} \frac{\del^3}{\del x_i\del x_j\del x_k}
\du v_iv_j v_k v_l F_{CE}\ud\delt^3+\ldots
\end{multline*}
While the second order moments yield exactly the fluxes of momentum, the
third order moments give rise to some additional terms. Using \eqref{mom3}, 
\begin{equation*}
\oh \frac{\del^2}{\del x_i\del x_j}\du v_iv_j v_l F_{CE}\ud\delt^2=
 \Biggl(\oh\lambda\Delta(\rho u_l)+\lambda\fr{\del}{\del x_l}\diverg(\rho\bfu)\Biggr)\delt.
\end{equation*}
According to \eqref{mom4-5}, the fourth order moment leads to
\begin{equation*}
-\frac{1}{6} \frac{\del^3}{\del x_i\del x_j\del x_k}
\du v_iv_j v_k v_l F_{CE}\ud\delt^3=-\oh\frac{\del}{\del x_l}\Delta\rho\delt+\ord\left(\frac{\delt^3}{\Ma^2}\right)
\end{equation*}
and fifth order moments are negligible since
\begin{equation*}
\du v_iv_j v_k v_l v_m F_{CE}\ud\delt^4=\ord\left(\frac{\delt^4}{\Ma^4}\right)=\ord(\delt^2).
\end{equation*}
Thus
\begin{multline}\label{tmomentum}
\rho^1 u_l^1=\rho u_l+
\Biggl(\oh\lambda\Delta(\rho u_l)+\lambda\fr{\del}{\del x_l}\diverg(\rho\bfu)-\oh\frac{\del}{\del x_l}\Delta\rho\\
-\frac{\del }{\del x_i}(\rho u_lu_i)-\fr{1}{\Ma^2}\frac{\del \rho }{\del x_l}+\fr{1}{\Rey}\frac{\del }{\del x_i}\left(2\rho S_{il}\right)+\fr{1}{\Rey}\frac{\del }{\del x_l}\left(\rho\frac{\del u_i}{\del x_i}\right) \Biggr)\delt +\ord(\delt^2).
\end{multline}
Since we assume that all appearing quantities are scaled, the equation for $\rho^1\bfu^1$ can only be balanced if $\del\rho/\del x_l=\ord(\Ma^2)$.
Hence, we assume that $\rho=\bar\rho(1+\Ma^2 p)$ for some constant $\bar\rho>0$ and a function $p$ which is assumed to be
of order one together with its derivatives in the limit under consideration.
(This is the standard scaling  of the density in isothermal, low Mach number flows.)
Using the additional information on $\rho$ and observing that $\Ma^2=\ord(\delt)$, we can simplify \eqref{tdensity} in lowest order to 
\begin{equation}\label{div0}
\diverg\bfu =\ord(\delt).
\end{equation}
This equation has to be understood in the sense that the order one assumption
on $\bfu$ and $p$ is only consistent if the divergence of $\bfu$ is $\ord(\delt)$.
Before we explain how the Kinetic Scheme guarantees the approximate divergence-free condition,
we note that relation \eqref{div0} and the structure of $\rho$ reduces \eqref{tmomentum} to the Navier Stokes equation with a first order error term 
\begin{equation}\label{NSi2}
\frac{\partial\bfu}{\partial t}+(\bfu .\nabla)\bfu+\nabla p=\left(\fr{1}{\Rey}+\oh\lambda\right)\Delta\bfu+\ord(\delt).
\end{equation}
To explain the mechanism that leads to \eqref{div0}, we use \eqref{tdensity} again, keeping the first order terms. After division
by $\delt$ and $\Ma^2$ this leads to 
\begin{equation*}
\frac{\del p}{\del t}+\diverg(p\bfu) +\frac{1}{\Ma^2}\diverg\bfu =\oh\lambda\Delta p+\ord(\fr{\delt^2}{\delt\Ma^2}).
\end{equation*}
To resolve the additional terms of order one on the right hand side, we 
have to expand \eqref{himo} one order higher. 
Using the explicit knowledge of the relevant moments, relation \eqref{div0} and our assumption on $\rho$, we find
\begin{equation*}
\frac{\del p}{\del t}+(\bfu\cdot\nabla)p +\frac{1}{\Ma^2}\diverg\bfu =\oh\lambda\Delta p+\diverg\left((\bfu\cdot\nabla)\bfu\right) +\ord(\delt).
\end{equation*}
Note that equations of this type are used in {\em pseudo-compressibility methods}
\cite{Hirsch, Rannacher} to ensure the divergence free condition.
In fact, it uses elements of Chorin's {\em artificial compressibility method} \cite{Chorin}
to replace $\diverg\bfu$ by the equation
\begin{equation*}
\eps\frac{\del p}{\del t}+\diverg\bfu=0
\end{equation*}
and of the pressure stabilization method
\begin{equation*}
\diverg\bfu-\eps\Delta p=0
\end{equation*}
which was originally used by Hughes, et. al., \cite{Hughes}.
However, the convection term and the nonlinear term which follow 
automatically from the kinetic approach are usually not considered.
We thus conclude that:
\begin{quote}
\em In the coupled limit $\delt,\Ma\to0$ with $\delt/\Ma^2=\lambda$ with the
assumption that $\rho=\bar\rho(1+\Ma^2 p)$ and that $\bfu, p$ and their derivatives
are order one functions, the Kinetic Scheme is consistent to the 
incompressible Navier Stokes equation with effective Reynolds number
\begin{equation*}
\fr{1}{\Rey'}=\fr{1}{\Rey}+\fr{\lambda}{2}.
\end{equation*}
The scheme can be viewed as a new pseudo-compressibility method.
\end{quote}
Note that in the case $\Rey=\infty$, the 
Chapman Enskog distribution reduces to a Maxwellian
and the Kinetic Scheme is equivalent to the Lattice Boltzmann method 
with relaxation parameter $t_R=\delt$.
Therefore, LBM can also be viewed as a pseudo-compressibility method in that case.
Since an additional viscosity term appears in the coupled limit $\delt,\Ma\to0$,
the Kinetic Scheme with $\nu=0$
still approximates the solution of an incompressible Navier Stokes equation.
As already mentioned earlier, this idea has been used in \cite{Michael3} to construct 
Navier Stokes solutions with a Kinetic Scheme which is just based on a discrete
Maxwellian. 

\section{Numerical Results and Discussions} \label{s5}
We first note that the term involving $\diverg\bfu$ in the Chapman Enskog distribution \eqref{discretefce1} 
is actually not important in low Mach number situations and thus can be neglected.
Note that such modifications
are very simple in the framework of Kinetic Schemes: by adding or deleting terms in the
distribution function, the macroscopic equations can easily be modified. In the case of
LBM, on the other hand, the Chapman Enskog distribution is implicitly given through
properties of the collision operator which makes it much harder to develop such schemes
for modifications of the incompressible Navier Stokes equation.

We adjust the viscosity parameter $\nu$ in the Chapman Enskog such that the effective viscosity
turns into the required one.
This prevents the numerical viscosity from spoiling the results of the simulations.
Altogether, we base our Kinetic Scheme on the following distribution function
(which is now written again in unscaled variables)
\begin{multline*}
F_{CE,i}(\rho,{\bf u}) = F_{i}^{*}\rho \Biggl( 1 +\frac{{\bf u} \cdot {\bf v}_{i}}{c_s^2}-\frac{1}{c_s^2}|\bfu|^2+\frac{1}{c_s^4}({\bf u} \cdot {\bf v}_{i})^2
-\left(\frac{\nu}{c_s^2}-\fr{\delt}{2}\right)S:\frac{{\bf v}_i\otimes{\bf v}_i}{c_s^2}\Biggr).
\end{multline*}
In a first test case, we apply the scheme to a Poiseuille flow in an infinitely long channel 
(in $x_1$--direction) of width one with a constant acceleration $g$.
The incompressible Navier Stokes solution for this case is explicitly known to be 
\begin{equation}\label{sol1}
u_1(x_2)=\fr{g}{2\nu}(1-x_2)x_2,\qquad u_2\equiv0
\end{equation}
with a constant pressure. 
In our simulation we choose $\rho\equiv 1$ initially. The infinitely long channel 
is modeled by periodic boundary conditions in $x_1$--direction. The fixed wall
conditions for $\bfu$ are enforced simply by setting $\bfu={\bf 0}$ at the boundary nodes.
In contrast to LBM, where the no--slip condition has to be enforced by properly setting
the incoming occupation numbers, no such complications are found here, because the
unknowns in the Kinetic Scheme are directly the flow variables $\rho$ and $\bfu$.
The boundary conditions for density can be obtained from the Navier Stokes
equation \eqref{NS1}. Multiplying the equation with the outward unit normal vector
$\bfn$ and observing that $\bfu={\bf 0}$ at the boundary, we get the condition
\begin{equation}\label{bc1}
c_s^2\fr{\del\rho}{\del \bfn}=\bfn\cdot\diverg\eta.
\end{equation}
For the exact solution \eqref{sol1} one easily checks that
\begin{equation*}
\diverg\eta=-\nu\rho g\begin{pmatrix} 1 \\ 0\end{pmatrix}
\end{equation*}
so that $\bfn\cdot\diverg\eta=0$ at the upper and lower walls giving rise to homogeneous boundary conditions for $\rho$.
(According to \cite{Hirsch}, 
homogeneous Neumann conditions for $\rho$ are also reasonable in more general, 
moderate flow situations.)
The force term is incorporated into our scheme by a splitting approach: in a first step
the Kinetic Scheme approximates the Navier Stokes evolution and in a second step,
the acceleration is taken care of by an explicit Euler step for the velocity variable.
To calculate the stress tensor $S$, we use central differences.

>From the solution \eqref{sol1}, we can see that the maximum velocity 
\begin{equation*}
U=\frac{g}{8\nu}
\end{equation*}
is obtained at the center of the channel.
By setting $g=0.01$ and $\nu=0.01$, we get $U=0.125$ (note that $U$ is the Mach number since $c_s=1$). With 11 points across the channel
and initial velocity $\bfu={\bf 0}$,
we find a numerical approximation which reproduces the predicted parabolic shape
(see Fig. \ref{f1}). The other velocity component stays zero and the density remains constant.
\begin{center}
\epsfig{file=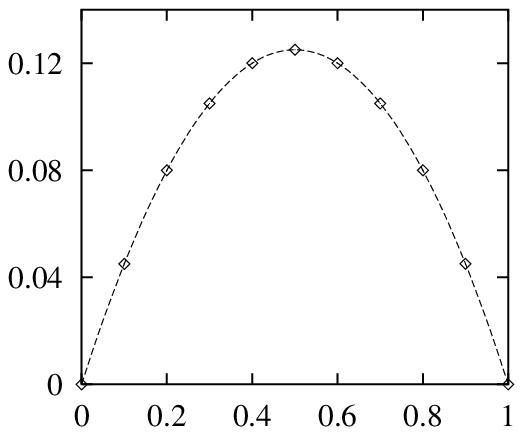,width=8cm}\\
\Caption{f1}{Poiseuille velocity profile}
\end{center}
Due to the symmetries in this simple test case, the incompressibility condition is satisfied
exactly. Consequently, compressibility errors are not present and the accuracy of the scheme
just depends on how closely the steady state is approximated. 
For several values of $t$, the $\EL^\infty$ error behaves
as depicted in Fig. \ref{f2}. In all calculations, the number of grid points is 11 and $U=0.125$.
\begin{center}
\epsfig{file=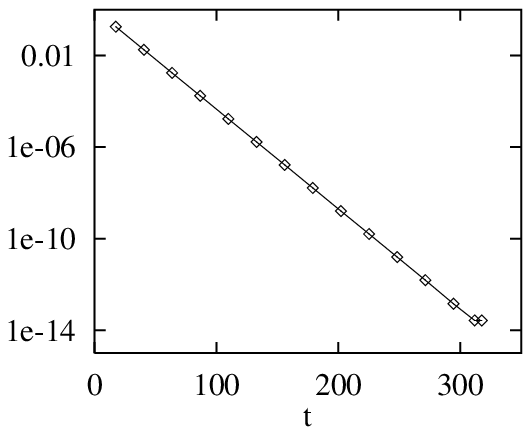,width=8cm}\\
\Caption{f2}{$\EL^\infty$--error versus time}
\end{center}
The next test case is a slight modification of the previous one, where the top wall
of the channel now moves with a fixed velocity $w$ in $x_1$--direction (Couette flow). 
Here, the exact solution differs from \eqref{sol1} by an additional linear term
\begin{equation}
u_1(x_2)=\fr{g}{2\nu}(1-x_2)x_2+wx_2,\qquad u_2\equiv0.
\end{equation}
Again, the $\bfu$--boundary condition at the moving wall is easily enforced by setting
$u_1=w$ and $u_2=0$. Using the same settings as above with $w=0.12$ the 
results are again 
in agreement with the exact solution (see Fig. \ref{f3}).
\begin{center}
\epsfig{file=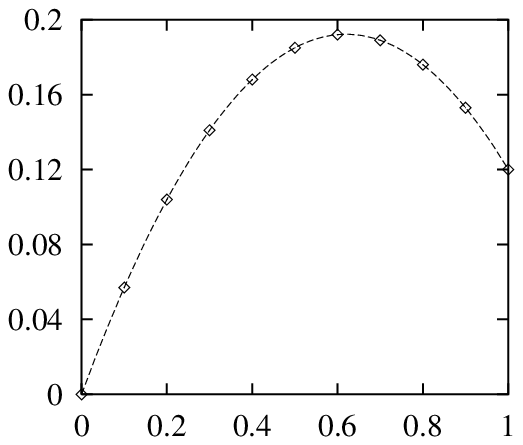,width=8cm}\\
\Caption{f3}{Velocity profile for Couette flow}
\end{center}
Our final test case is the driven cavity flow problem. The 
incompressible fluid is now bounded by a square enclosure with side length one.
The top lid, which moves with velocity $U$, generates the fluid motion in the cavity
which shows typical vortex phenomena. 
For our calculations we use a $129\times 129$ uniform grid and Reynolds numbers 100, 400 and 1000.
The lid velocity $U$ is set to one and $c_s=10$ in all cases.
The calculations are initialized with $\rho=1$ and $\bfu={\bf 0}$ inside the cavity.
As termination criterion we choose a residue fall of 3.75 decades in the equation for $\rho$.
The typical number of cycles to get steady state solutions is 100,000. We remark that no
special attention has been paid to acceleration of convergence. Our aim is only to show
that the new discrete velocity method works for complex test problems.
(Note that the pressure develops singularities in the top corners due to the jump in the
boundary conditions for velocity.)
In order to demonstrate that the Kinetic Scheme can be used like a Lattice Boltzmann method
we implement the boundary conditions using the fast {\em bounce back} algorithm \cite{CHL91}.
To explain this approach we remark that on the kinetic level, boundary conditions are required
for the transport part of the equation
\begin{equation}\label{tra}
\fr{\del f}{\del t}+\bfv.\nabla f=0.
\end{equation}
Since \eqref{tra} is a linear hyperbolic equation, information has to be provided for those
characteristics which enter the domain at a boundary. In our model, the
characteristics are straight lines along the discrete velocity directions $\bfv_1\dots\bfv_8$.
The bounce back condition sets the value for the information of an incoming direction equal to the information
that leaves the domain in the opposite direction, which is easily available due to the
symmetry of the discrete velocity set.
It can be shown \cite{CHL91} that these conditions simulate no slip conditions
at the Navier Stokes level. At the upper lid, a modification is required which takes care
of the momentum flux generated by the movement \cite{Ladd94}.
To illustrate our results, the horizontal velocity component $u_1$ is shown along a vertical
section through the center of the cavity (Fig. \ref{f4}). Similarly, we plot the
vertical component $u_2$ along the central horizontal section (Fig. \ref{f5}).
The results are compared with those obtained by Ghia et. al. \cite{GGS82} and they 
are in good agreement.
In Figures \ref{f4} and \ref{f5}, the symbols refer to the tabulated simulation results in \cite{GGS82} and the lines refer to the results obtained by the new
Kinetic Scheme.
\begin{center}
\epsfig{file=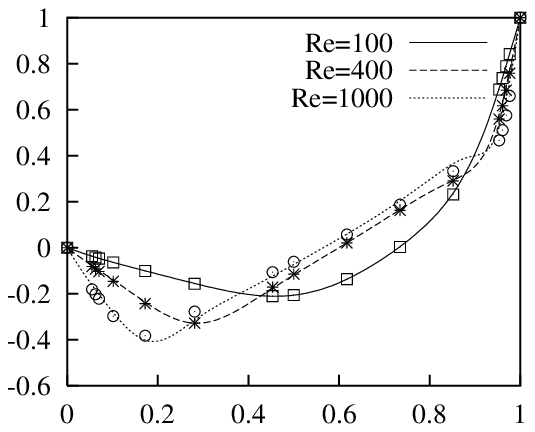,width=10cm}\\
\Caption{f4}{$u_1$-velocity along a vertical line through the center of the cavity}
\end{center}
\begin{center}
\epsfig{file=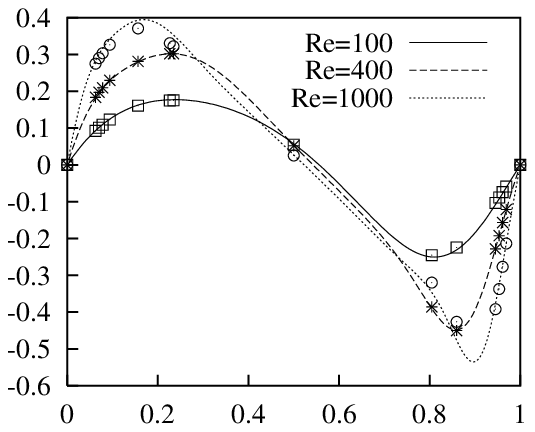,width=10cm}\\
\Caption{f5}{$u_2$-velocity along a horizontal line through the center of the cavity}
\end{center}
Plots of the stream functions are given in Figures \ref{f6} to \ref{f8}.
We remark that the stream function $\psi$ is, strictly speaking, not well defined because
the approximate velocity field is not exactly divergence free.
In \cite{HouZou}, this problem is discussed for the Lattice Boltzmann method and we 
use the proposed numerical procedure for the calculation of $\psi$ (integration of $u_2$ along horizontal sections
from left to right). The levels of the isolines are those from \cite{GGS82}. We limit ourselves in this study to the use of
uniform grids, as our purpose is to show that the new discrete velocity model works. With clustered
grids, the solution can be different \cite{SND98}.
\begin{center}
\begin{tabular}{cc}
\epsfig{file=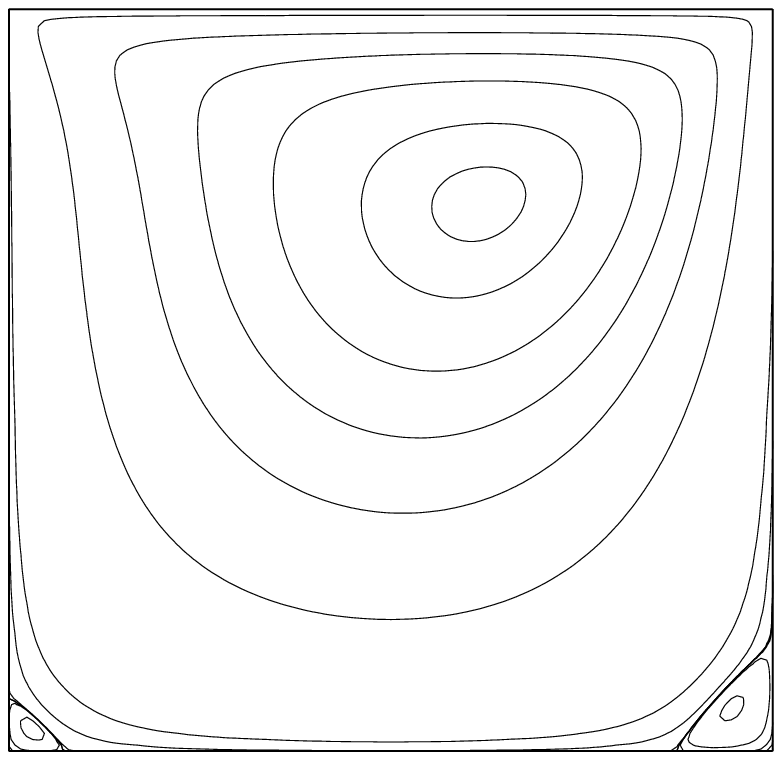,width=6cm} & \epsfig{file=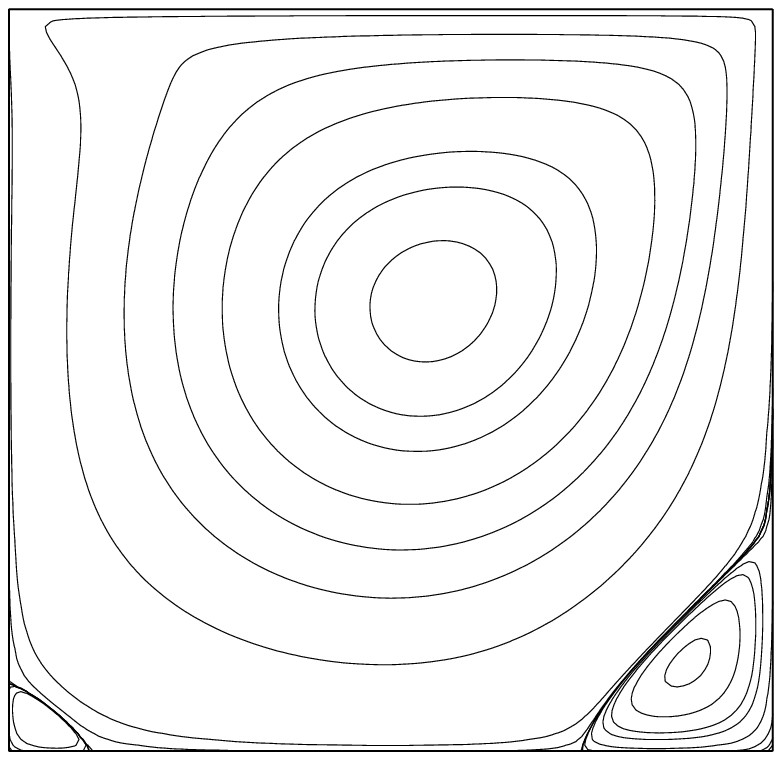,width=6cm}\\
\Caption{f6}{$\Rey=100$} & 
\Caption{f7}{$\Rey=400$}  
\end{tabular}
\end{center}
\begin{center}
\epsfig{file=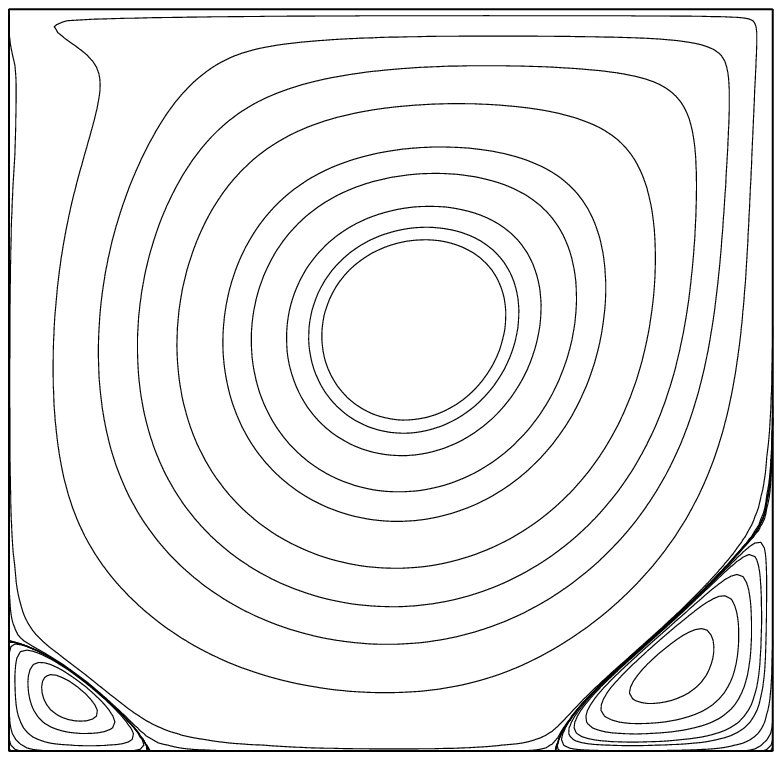,width=6cm}\\
\Caption{f8}{$\Rey=1000$}
\end{center}
The numerical cost of the Kinetic Scheme is directly comparable to that of the Lattice Boltzmann
method based on the D2Q9 model. Both algorithms have the same structure consisting of a propagation and a collision step.
The only difference is that in the Kinetic Scheme the stress tensor has to be calculated (by taking central
differences of the velocity field) and that the equilibrium distribution is extended by the viscosity term.
On the other hand, the Kinetic Scheme needs less memory because there is no need to store the occupation numbers.
Apart from two copies of $\rho$ and $\bfu$ (new and old time step) an efficient implementation requires three more variables
to store the stress tensor. Altogether a 2D computation needs nine floating point variables per node, independent of the
underlying discrete velocity model. Compared to that, D2Q9 Lattice Boltzmann methods need 21 variables per node
(for $\rho,\bfu$ and two copies of the occupation numbers $f_0,\dots,f_8$) and the number increases if models with more velocities are
used. Also, when passing over to 3D calculations, the memory usage of the Kinetic scheme increases by five variables per site whereas
D3Q15 Lattice Boltzmann methods need 13 more variables in each node. Of course, the discrepancy becomes even larger if multi-phase flows
are simulated. Then, for simple algorithms, the memory requirement has essentially to be multiplied by the number of participating species.
Taking these considerations into account, Kinetic Schemes seem to be a powerful alternative to Lattice Boltzmann methods.
On the one hand, they are formulated in the same kinetic framework allowing the use of LBM specific solution techniques
(kinetic boundary conditions, treatment of phase boundaries in multi-phase flows, etc.). On the other hand, Kinetic Schemes
only use the actual flow variables and thus can profit directly from established Finite Difference or Finite Volume methods.
In addition, the memory consumption is greatly reduced compared to Lattice Boltzmann algorithms.

\vspace{0.5\baselineskip}

Since the idea of LBM is to use
a kinetic model which is as simple as possible under the constraint that the macroscopic limit equations are correct,
the method is not capable 
of quantitatively predicting the behavior of a rarefied gas 
and should therefore only be applied close to equilibrium situations. To show consistency of LBM to the Navier Stokes equation,
exactly this equilibrium assumption is used in the {\em Chapman Enskog} expansion which amounts to
assuming that the occupation numbers are given by a Chapman Enskog distribution.
A natural idea is therefore to build the Chapman Enskog distribution directly into the algorithm which is
exactly the construction principle of the present Kinetic Scheme.
Thus, Kinetic Schemes can be viewed as a consequent advancement of the Lattice Boltzmann Method.

\vspace{0.5\baselineskip}

We conclude our discussion with a remark concerning the extension to the full Navier Stokes system including
the energy equation. A fundamental problem of the basic Lattice Boltzmann method based on a simple BGK collision operator
is that the Prandtl number is not a free parameter. In a Kinetic Scheme, the heat conduction and viscosity parameters
enter directly into the Chapman Enskog distribution (similar to the continuous case \eqref{fCEdefs}) and thus can naturally
be varied independently.

\section{Conclusions} 
     The similarities and differences between the Lattice Boltzmann Method, 
which has recently become popular, and the Kinetic Schemes, which are 
routinely used in Computational Fluid Dynamics, are studied.  A new 
discrete velocity model for the numerical simulation of incompressible 
Navier--Stokes equations is presented by combining both the approaches.  
This approach of Kinetic Schemes with discrete distributions is shown 
to be more convenient and useful compared to the Lattice Boltzmann 
Method. Since both methods coincide for a particular choice of parameters,
the analysis of the Kinetic Scheme also applies directly to LBM in that
case. In particular, the conclusion that the Kinetic Scheme is a
special pseudo-compressibility method illuminates the Lattice Boltzmann
approach.

\end{document}